\documentclass[final,5p,times,twocolumn,authoryear]{elsarticle}
\usepackage[normalem]{ulem}
\usepackage[utf8]{inputenc}
\usepackage[T1]{fontenc}
\usepackage{dcolumn}
\usepackage{bm}
\usepackage{color}
\usepackage{amssymb}
\usepackage{amsmath}
\usepackage{graphicx}
\usepackage{amsfonts}
\usepackage{slashed}
\usepackage{pstricks}
\usepackage{float}
\usepackage{multirow}
\usepackage{hyperref}
\usepackage{array}
\allowdisplaybreaks
\usepackage{dcolumn}
\usepackage{epsf}
\usepackage{float}
\usepackage{subcaption}
\usepackage[nice]{nicefrac}

\begin{document}
\title{Modeling inclusive electron-nucleus scattering with Bayesian artificial neural networks}

\author{Joanna E. Sobczyk}
\affiliation{organization={Institut f\"ur Kernphysik and PRISMA$^+$ Cluster of Excellence, Johannes Gutenberg-Universit\"at}, postcode={55128},city={Mainz}, state={Germany}}

\author{Noemi Rocco}
\affiliation{Theoretical Physics Department, Fermi National Accelerator Laboratory, P.O. Box 500, Batavia, Illinois 60510, USA}

\author{Alessandro Lovato}
\affiliation{Physics Division, Argonne National Laboratory, Argonne, IL 60439}
\affiliation{INFN-TIFPA Trento Institute of Fundamental Physics and Applications, 38123 Trento, Italy}

\date{\today}
\begin{abstract}
We introduce a Bayesian protocol based on artificial neural networks that is suitable for modeling inclusive electron-nucleus scattering on a variety of nuclear targets with quantified uncertainties. Unlike previous applications in the field, which directly parameterize the cross sections, our approach employs artificial neural networks to represent the longitudinal and transverse response functions. In contrast to cross sections, which depend on the incoming energy, scattering angle, and energy transfer, the response functions are determined solely by the energy and momentum transfer to the system, allowing the angular component to be treated analytically. We assess the accuracy and predictive power of our framework against the extensive data in the quasielastic inclusive electron-scattering database. Additionally, we present novel extractions of the longitudinal and transverse response functions and compare them with previous experimental analysis and nuclear ab-initio calculations.
\end{abstract} 

\maketitle

\section{Introduction}
Electron scattering has long been recognized as an excellent tool for probing nuclear structure, including testing the limits of the nuclear shell model~\citep{Benhar:2004hm}. Due to the perturbative nature of electromagnetic interactions, electrons offer several advantages over hadrons, providing valuable high-precision cross-sections. Currently, numerous datasets are available, covering various nuclear targets and kinematics, which are sensitive to a rich array of underlying nuclear dynamics~\citep{Benhar:2006er}. The field has recently experienced renewed interest, motivated by its interplay with high-precision measurements of neutrinos and their oscillations~\citep{Amaro:2019zos,CLAS:2021neh,Ankowski:2022thw}.  

The inclusive electron-nucleus scattering cross section is characterized by a variety of reaction mechanisms, whose relative importance strongly depends on the energy transfer to the nucleus~\citep{Benhar:2006wy}. The low-energy region features narrow peaks corresponding to transitions to excited states, which are determined by the structure details of the specific nucleus. At slightly higher energies, broader peaks associated with collective modes can also appear. For energies on the order of hundreds of MeV, the leading mechanism is quasielastic scattering, where the probe primarily interacts with individual bound nucleons. Corrections to this mechanism arise from processes in which the lepton couples to interacting nucleons via nuclear correlations and two-body currents. At higher energy transfers, the strength is driven by pion production, involving mechanisms such as the excitation of baryonic resonances. The deep inelastic scattering region corresponds to even higher energies, where the probe can resolve the internal quark and gluon structure of the nucleons. 

Devising a framework that is flexible enough to accurately model these distinct regions, where various degrees of freedom are at play, is exceedingly complex. The availability of such a framework, besides being valuable in its own right, can significantly impact neutrino-oscillation experiments, as the neutral- and charge-current processes share the same nuclear dynamics of the target state and the vector part of the electroweak current~\citep{Benhar:2015wva,NuSTEC:2017hzk,Amaro:2019zos}.

The inclusive electron-nucleus cross section can be expressed in terms of the longitudinal and transverse response functions, which correspond to transitions induced by a polarized virtual photon. This separation offers insights into nuclear dynamics, with the response functions conveniently expressed in terms of nuclear transition amplitudes. Several experiments have performed Rosenbluth separation in the past~\citep{Meziani:1984is, Dytman:1988fi, Zghiche:1993xg, Jourdan:1996np, Williamson:1997zz,  Morgenstern:2001jt, Bodek:2022gli}, and their data has been used to validate ab initio approaches, including Green's Function Monte Carlo (GFMC) and Coupled Cluster (CC).

The GFMC is a continuum quantum Monte Carlo method that solves the quantum many-body problem with percent-level accuracy, fully capturing the complexity of long- and short-range correlations~\citep{Carlson:2014vla}. It has been used to obtain accurate calculations of inclusive electroweak response functions~\citep{Lovato:2016gkq, Lovato:2020kba} for $^4$He and $^{12}$C, including both one- and two-body current contributions, initial state correlations and final-state interactions. The response functions are reconstructed from the imaginary-time propagators by approximately inverting the Laplace transform~\citep{Lovato:2015qka, Raghavan:2020bze}. However, the exponentially growing computational cost with the number of nucleons limits the applicability of GFMC to nuclei up to $^{12}$C.

The CC approach includes correlation effects onto a starting Slater determinant as $n$-particle $n$-hole excitations based on an exponential ansatz~\citep{hagen2014}. Due to the polynomial scaling with number of nucleons, it is particularly suited for medium-mass and even heavy systems~\cite{Hu:2021trw}.  Response functions can be computed by combining CC with the Lorentz integral transform (LIT) method~\citep{efros1994}, originally proposed in~\cite{Bacca:2013dma}. The LIT-CC technique was used to obtain the longitudinal and transverse response functions of $^4$He and $^{40}$C, employing chiral effective field theory potentials and one-body currents~\citep{Sobczyk:2021dwm, Sobczyk:2023sxh}. Recently, a strategy to extend CC calculations to open-shell nuclei by using the equation-of-motion CC framework and adding two nucleons to a closed-shell system has been presented in~\cite{Bonaiti:2024fft}.

Both GFMC and CC approaches have limitations: they provide only fully inclusive reactions and are restricted to non-relativistic kinematics, although some relativistic corrections have been included in GFMC~\citep{Rocco:2018tes, Nikolakopoulos:2023zse}. Moreover, calculating responses in both cases requires inverting integral transforms, a procedure that entails nontrivial difficulties and can lead to large systematic errors~\citep{Sobczyk:2021ejs, Raghavan:2020bze}.

In this study, we put forward an artificial neural network (ANN) representation of the longitudinal and transverse response functions of selected nuclear targets, trained on inclusive electron scattering data. Our approach differs substantially from recent works that focused on directly predicting the $(e,e')$ inclusive scattering cross-sections~\citep{AlHammal:2023svo, Kowal:2023dcq}. Not only are we able to predict cross sections, but more importantly, we extract valuable information on the response functions, providing insights into nuclear systems that can be used to benchmark theoretical studies. We validate our findings against available Rosenbluth separation data. Furthermore, we address several nuclei, allowing us to make predictions for systems with very scarce experimental data available (e.g., $^6$Li, $^{16}$O), where Rosenbluth separation would not otherwise be possible.
To assess the uncertainty in our predictions and mitigate the overfitting problem, we adopt a Bayesian approach, representing the weights and biases of the network as probability distributions rather than fixed values~\citep{Neal:2012}. By treating the network parameters probabilistically, we account for the inherent uncertainty in the training data and the model itself. The predictions we present are sampled from a posterior distribution obtained through Markov Chain Monte Carlo (MCMC) methods. Hence, we not only provide point estimates of the response functions but also quantify the uncertainty associated with these predictions, reflecting the model's reliability and providing confidence intervals for its outputs.

This work is organized as follows: In Sec.\ref{sec:methods}, we present the methodology employed to construct our ANN architecture and the training algorithms. Sec.\ref{sec:results} is dedicated to the results obtained for both cross sections and response functions. In Sec.~\ref{sec:conclusions}, we draw our conclusions and outline future perspectives of our work.

\section{Methodology}
\label{sec:methods}
The double differential cross section of the inclusive electron-nucleus scattering process in which an electron of initial four-momentum $k=(E, {\bf k})$ scatters off a nuclear target at rest to a state of four-momentum $k^\prime=(E^\prime, {\bf k}^\prime)$
can be written as  
\begin{align}
\left(\frac{d^2\sigma}{d E^\prime d\Omega^\prime}\right)_e & =\frac{\alpha^2}{Q^4}\frac{E^\prime}{E}L_{\mu\nu}R^{\mu\nu}\, ,
\label{xsec:em1}
\end{align}
where $\alpha\simeq1/137$ is the fine structure constant, 
and $\Omega^\prime$ is the scattering solid angle in the direction specified by ${\bf k}^\prime$.
The energy and the momentum transfer are denoted by $\omega$  and {\bf q}, respectively, with $Q^2=-q^2={\bf q}^2-\omega^2$.
The lepton tensor $L_{\mu\nu}$ is fully specified by the measured electron kinematic variables; its expression can be found in several articles, see for example~\cite{Andreoli:2021cxo}. 
The hadronic tensor describes the transition between the initial and final hadronic states $|\Psi_0\rangle$ and 
$|\Psi_f\rangle$ with energies $E_0$ and $E_f$
\begin{align}
R^{\mu\nu}({\bf q},\omega)&= \sum_f \langle \Psi_0|J^{\mu \, \dagger}({\bf q},\omega)|\Psi_f\rangle \langle \Psi_f| J^\nu({\bf q},\omega) |\Psi_0 \rangle \nonumber\\
&\times\delta (E_0+\omega -E_f)\, .
\label{eq:had_tens}
\end{align}
The sum includes all the possible final states, both bound and in the continuum, and $J^\mu({\bf q},\omega)$ is the nuclear current operator. 
For inclusive processes, the cross section of Eq.~\eqref{xsec:em1} only depends on the longitudinal and transverse response functions,
$R_L({\bf q}, \omega) \equiv R^{00}({\bf q}, \omega)$ and $R_T({\bf q},\omega)\equiv [R^{xx}({\bf q}, \omega) + R^{yy}({\bf q}, \omega)] / 2$, respectively
\begin{align}
\left(\frac{d^2\sigma}{d E^\prime d\Omega^\prime }\right)_e &  =\left( \frac{d \sigma}{d\Omega^\prime} \right)_{\rm{M}} \left[ \frac{q^4}{{\bf q}^4}  R_L({\bf q},\omega) \right. \nonumber\\
&\left. + \left(\tan^2\frac{\theta}{2}-\frac{1}{2}\frac{q^2}{{\bf q}^2}\right)  R_T({\bf q},\omega) \right] \ .
\label{eq:x:sec}
\end{align}
The Mott cross section 
\begin{align}
\label{Mott}
\left( \frac{d \sigma}{d \Omega^\prime} \right)_{\rm{M}}= \left[ \frac{\alpha \cos(\theta/2)}{2 E^\prime\sin^2(\theta/2) }\right]^2
\end{align} 
only depends upon the scattering angle $\theta$ and on the outgoing electron energy $E^\prime$. 

Experimentally, the procedure used to separate $R_L$ and $R_T$, known as Rosenbluth separation, consists on defining the quantity
\begin{align}
\Sigma({\bf q},\omega,\epsilon)&= \epsilon \, \frac{{\bf q}^4}{Q^4} \Big(\frac{d^2\sigma}{d E^\prime d\Omega^\prime }\Big)_e \Bigg/\left( \frac{d \sigma}{d\Omega^\prime}  \right)_{\rm{M}}  \nonumber\\
& = \epsilon R_L({\bf q},\omega) + \frac{1}{2} \frac{{\bf q}^2}{Q^2} R_T({\bf q},\omega)
\label{eq:Rosenbluth}
\end{align}
and plotting it as a function of the virtual photon polarization defined as
\begin{align}
    \epsilon= \left( 1 + \frac{2 {\bf q}^2}{Q^2} \tan^2 \frac{\theta}{2}\right)^{-1}\ .
\end{align}
As the scattering angle ranges between 180 to 0 degrees, $\epsilon$ varies between 0 and 1. Within this approach, $R_L$ is the slope while $({\bf q}^2/2 Q^2) R_T$ is the intercept of the linear fit to data. Note that Eq.~\eqref{eq:Rosenbluth} can only be applied if the Born approximation is valid and if the data have already been corrected to account for Coulomb distortions of the electron wave function.

In our calculations we assume that the Born approximations holds, and following~\cite{Aste:2007sa,Wallace:2008ev}, we account for the Coulomb distortion effects by using an effective momentum approximation. For target nuclei with a large number of protons, the  Coulomb field induces a distortion of the electron wave function yielding a modification in the $(e, e')$ cross section and inducing sizable effects in the longitudinal and transverse separation of the electromagnetic response.  Since a highly relativistic particle is moving nearly on a straight line inside a potential $V({\bf r})$, its momentum can be rewritten as 
$|{\bf k}^{(')}_{\rm eff}|=E^{(')}+V({\bf r})$ where we neglected the particle's mass. This expression is valid for the initial (final) electron momentum. For large momentum transfer, the knockout process is nearly local, therefore one can consider a potential value $\bar{V}$ which is obtained by taking the average over the nuclear density profile $\rho({\bf r})$. 
If we approximate the nucleus with a homogeneously charged sphere with radius
$R_{\rm sp}= (1.1 A^{1/3} + 0.86 A^{-1/3})$ and charge number $Z$, the electric potential in the center of the sphere is
given by $V(0)=-3\alpha$. It follows that the potential averaged over the volume of the sphere is 
$\bar{V} = 3/2  Z \alpha/R_{\rm sp}$.
The modulus of the effective momentum transfer is obtained as
\begin{equation}
    |{\bf q}_{\rm eff}| = \sqrt{|{\bf k}_{\rm eff}|^2 + |{\bf k^\prime}_{\rm eff}|^2 - 2 |{\bf k}_{\rm eff}||{\bf k^\prime}_{\rm eff}| \cos\theta}\,.
    \label{eq:eff_q}
\end{equation}
A focusing factor can be introduced to account for the attractive nucleus, focusing the electron wave function in the nuclear region. However, if the same $\bar{V}$ is used for both the effective momentum and the effective focusing factors, a cancellation of the focusing factors occurs. Therefore, in the effective momentum approximation, we simply replace $q$ with $q_{\rm eff}$ in the cross section expression of Eq.~\eqref{eq:x:sec}. This replacement accounts for both the momentum enhancement of the electron near the attractive nucleus and the focusing of the electron wave function.

\subsection{Neural network architecture}
\begin{figure}[!t]
\includegraphics[width=0.495\textwidth]{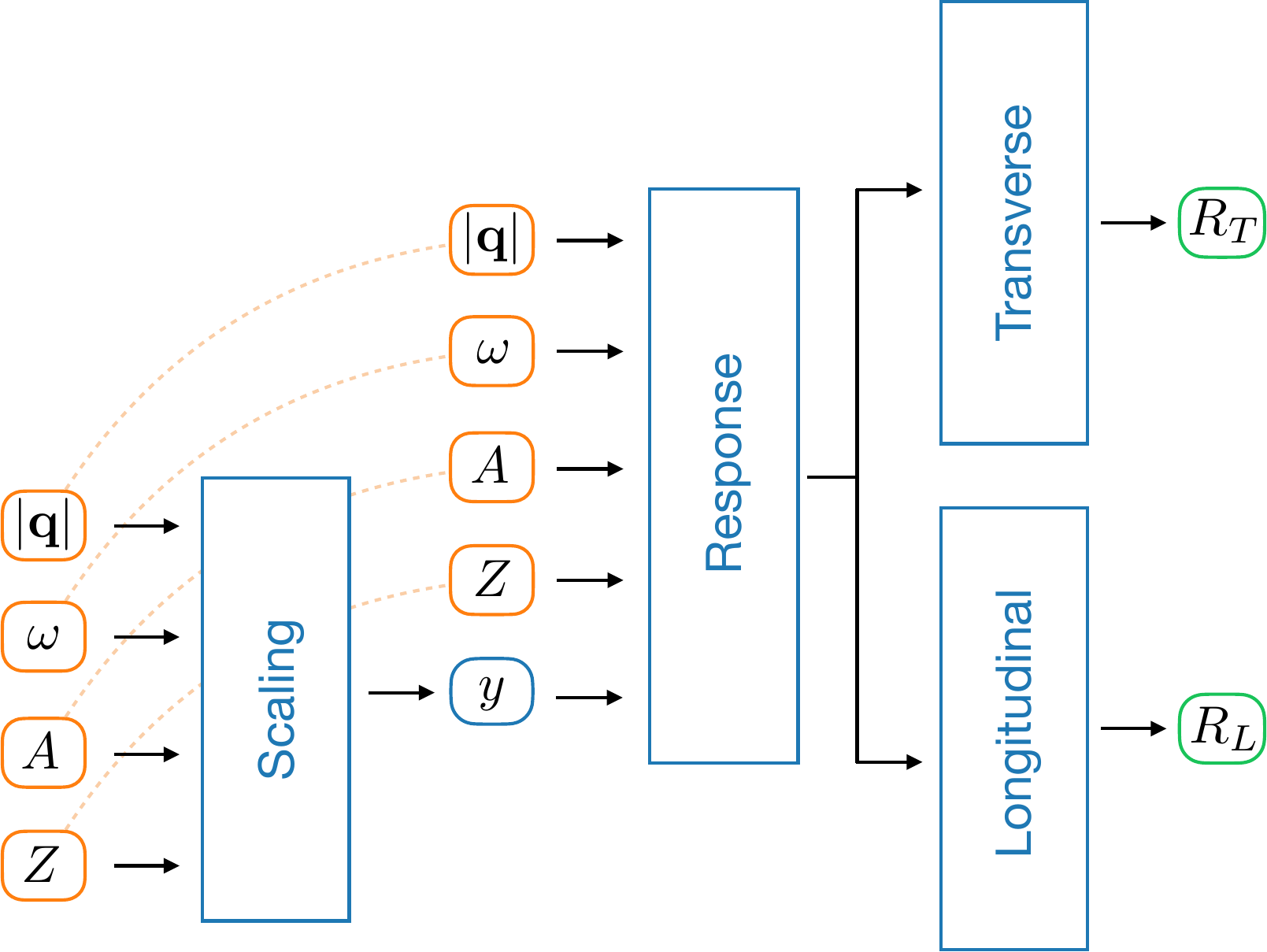}
  \caption{Schematic representation of the ANN architecture we employ to represent the electromagnetic longitudinal and transverse response functions. \label{fig:ANN}}
\end{figure}

The longitudinal and transverse electromagnetic response functions are outputs of the ANN architecture illustrated in Fig.~\ref{fig:ANN}. The input of the network is a four-dimensional array obtained by concatenating the energy and momentum transfer with the number of nucleons and the number of protons: $(\omega, |{\bf q}|, A, Z)$. The input energies are in GeV, ensuring that their maximum value is of the order of one. To mitigate scale differences among the inputs, which could cause certain features to dominate the learning process, we employ a standard score~\citep{Kreyszig:1979} to scale $Z$ as
\begin{equation}
Z_{\rm resc}= \frac{Z - Z_{\rm avg}}{\Delta Z}\,.
\end{equation}
In this equation, $ Z_{\rm avg}$ is the average value of $Z$, calculated as the sum of all $Z_{i=1,\ldots,N}$ values divided by $N$, the total number of nuclei analyzed. The term $\Delta Z$ is defined as the difference between the maximum and minimum values of $Z$ within the range of nuclei considered. The same normalization procedure is applied to the particle number $A$.

%The input energies are in GeV, so that their maximum is of the order of one. To reduce scale difference among the inputs, which may cause some features to dominate the learning process, we normalize $Z$ using the formula
%\begin{equation}
%Z_{\rm resc}= \frac{\Big(Z - Z_{\rm avg}\Big){\Delta Z}\,.
%\end{equation}
%In the above equation, $Z_{\rm avg}$ is the average value of $Z$, calculated as the sum of all $Z_{i=1,
%\ldots,N}$ values divided by $N$, the total number of nuclei analyzed, while $\Delta Z$ is defined as the difference between the maximum and minimum values of $Z$ within the range of nuclei considered. The same normalization procedure is adopted for the particle number $A$. 
Inspired by the scaling properties of electromagnetic responses~\cite{Day:1987az,Donnelly:1998xg,Benhar:1999ts}, we preprocess the input through a ``Scaling'' network, whose single output is the variable $y(\omega, |{\bf q}|, A, Z)$. Note, however, that since we do not pretrain the scaling network, $y$ does not necessarily correspond to the scaling variable commonly employed in the literature. Leveraging the concept of skip-connections~\cite{Srivastava:2015}, the output of the scaling network is concatenated with the other inputs, forming the five-dimensional array $(y, \omega, q , A,Z)$, which is then input to
a ``Response'' network. The latter produces a $32$-dimensional output which is then taken as input to both the ``Longitudinal'' and ``Transverse'' networks. These latter networks are completely independent and each provides a single output corresponding to the longitudinal and transverse responses, respectively. The Scaling, Response, Transverse, and Longitudinal networks are multilayer perceptrons (MLPs) with two hidden layers, each comprised of $32$ neurons and using the hyperbolic tangent activation function. To ensure positive definiteness, an exponential function is employed to transform the raw outputs of both the longitudinal and transverse MLPs and obtain $\hat{R}_L$ and $\hat{R}_T$. We collectively denote the weights and biases of the ANN with $\mathcal{W}={w_1, \dots, w{N}}$ --- there are a total of 6787 parameters. 

\subsection{Bayesian training}
The double differential cross section corresponding to a given nuclear species, incoming energy of the lepton, scattering angle, and energy transfer, dubbed $\hat{y}_i(\mathcal{W})$, is obtained plugging $\hat{R}_L$ and $\hat{R}_T$ into Eq.~\eqref{eq:x:sec} evaluated at the corresponding energy transfer, while the effective momentum transfer of Eq.~\eqref{eq:eff_q} accounts for Coulomb distortion effects.

\begin{figure*}[!htb]  
   \centering
    \includegraphics[width=\textwidth]{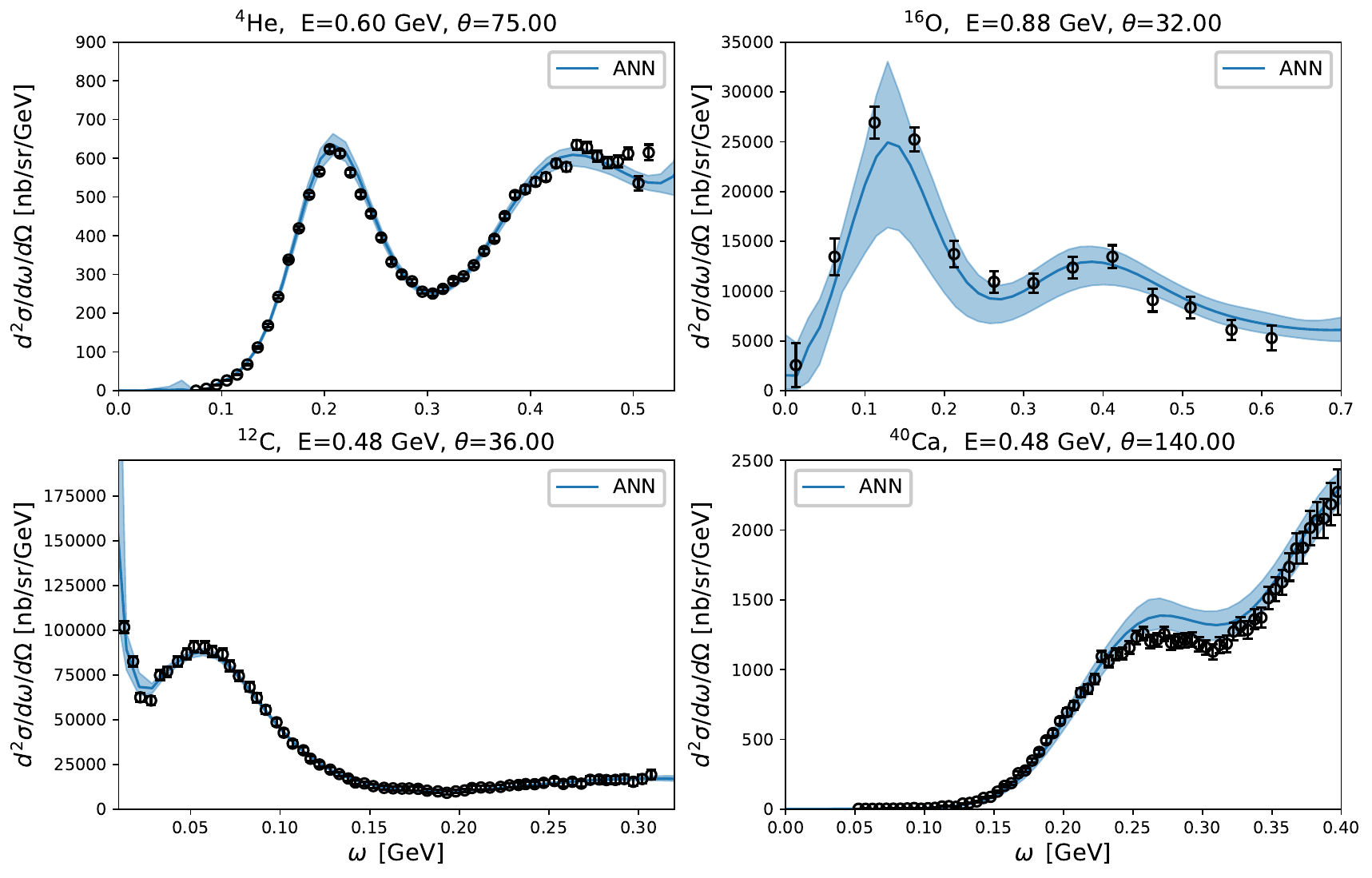}
    \caption{Results on test data for four symmetric nuclei. The uncertainty band encompasses the total spread of the ANN predictions. Experimental data taken from~\cite{Zghiche:1993xg, Barreau:1983ht, Anghinolfi:1996vm, Meziani:1984is}.}
 \label{fig:test_xsection}
\end{figure*}

We train our ANN using the quasielastic electron nucleus scattering archive of \cite{Benhar:2006er} on five selected light and medium-mass nuclei, all with an equal number of protons and neutrons: $^4$He, $^6$Li, ${12}$C, $^{16}$O and $^{40}$Ca. Following~\cite{Kowal:2023dcq}, we remove from our analysis the datasets on $^{12}$C from~\cite{Zeller:1973ge}. Based on our preliminary analysis they stay in tension with all other experiments. For $^{16}$O, we add to our analysis the data from  ~\cite{Anghinolfi:1996vm}, which are not included in quasielastic electron nucleus scattering archive of~\cite{Benhar:2006er}. 

A critical aspect of this work consists in quantifying the uncertainty associated with the ANN predictions. To this aim, we leverage Bayesian statistics and treat $\mathcal{W}$ as probability distributions~\citep{Neal:2012}. Using Bayes’ theorem, the posterior probability of the parameters $\mathcal{W}$ given the measured cross sections $Y$ can be written as
\begin{equation}
    P(\mathcal{W}|Y) = \frac{P(Y|\mathcal{W}) P(\mathcal{W})}{P(Y)}\, ,
\end{equation}
where $P(Y|\mathcal{W})$ is the likelihood and $P(\mathcal{W})$ is the prior density of the parameters~\citep{Utama:2015hva}. As in \cite{Neal:2012}, we assign a normal Gaussian prior for each neural network parameter
\begin{equation}
P(\mathcal{W}) = \frac{1}{(2\pi)^{N/2}}\exp\left(\sum_{i=1}^N - \frac{w_i^2}{2} \right)\,.
\end{equation}
Note that such prior corresponds to $l_2$ regularization with unit weight.

Following standard practice, we assume a Gaussian distribution for the likelihood based on a loss function obtained from a least-squares fit to the empirical data 
\begin{equation}
    P(Y|\mathcal{W}) = \exp\left(-\frac{\chi^2}{2}\right)\,,
\end{equation}
where 
\begin{align}
    &\chi^2 = \sum_{i=1}^N \frac{ \left[ y_i-\hat{y}_i(\mathcal{W}))\right]^2 }{ \sigma_i^2 }\,.
\end{align}
In the above equation, $y_i$ is the $i$-th experimental value of the cross section and the sum runs over the kinematics and nuclei included in the training dataset. We augment the experimental errors $\sigma_i$ listed in~\cite{Benhar:2006er} including an additional term proportional to the experimental cross section value: $\sigma_i \to \sigma_i + 0.05 y_i$. The primary reason behind this choice is that experimental errors are in general small and most experiments report an additional few-percent systematic uncertainty. 

All of our numerical simulations are performed using the JAX Python library~\cite{jax2018github}. The posterior distribution is sampled leveraging the NumPyro No-U-Turn Sampler extension of Hamiltonian Monte Carlo (HMC) ~\citep{phan2019composable,bingham2019pyro}. Additionally, we implemented the standard HMC algorithm as outlined in Ref.~\cite{hoffman2011nouturn} and found results that are consistent with those obtained using the NumPyro package.

\section{Results}
\label{sec:results}
In the first part of the analysis, we split experimental data into training and test datasets, containing $80\%$, and $20\%$ of the measured cross sections, respectively. Since the experimental errors at each kinematic setup for a given nucleus are likely to be correlated, we never split data coming from a single experiment. However, we have no way to account for correlated errors among different kinematics. 

\begin{figure}[t]
    \includegraphics[width=0.45\textwidth]{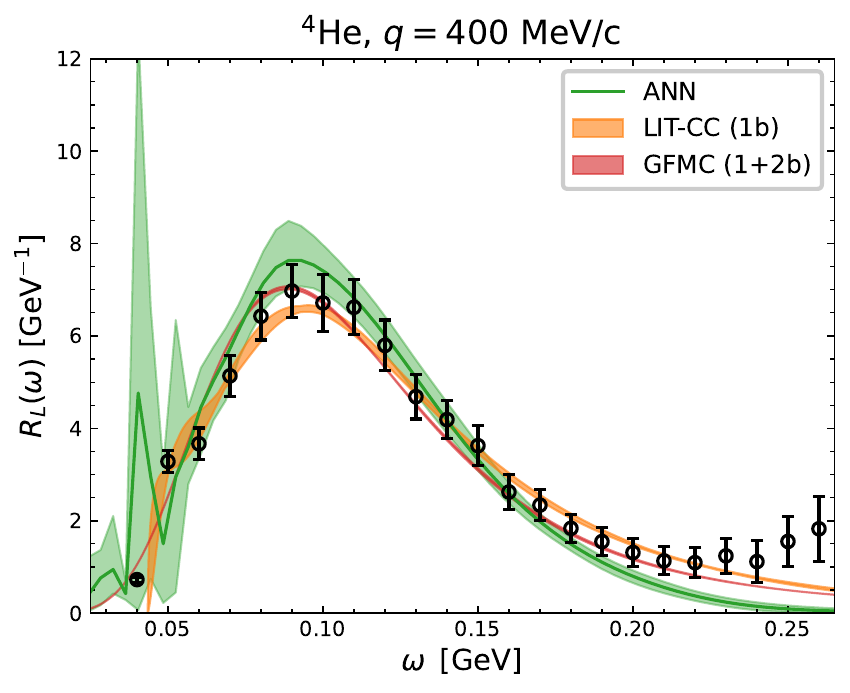}
\includegraphics[width=0.45\textwidth]{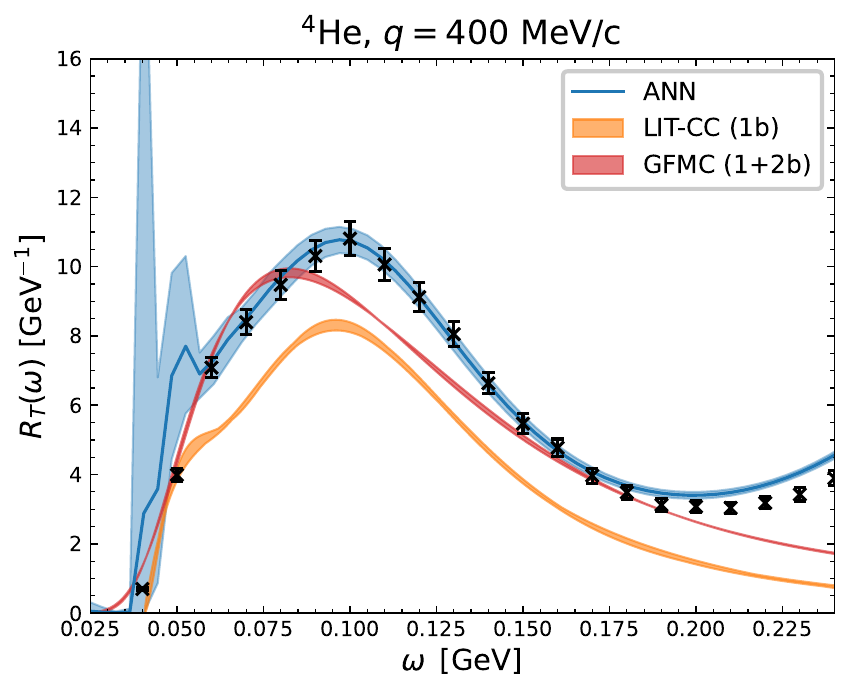}
  \caption{Electromagnetic longitudinal (upper panel) and transverse (lower panel) responses of $^4$He at $q=400$ MeV. The ANN results are compared with theoretical calculations~\citep{Lovato:2015qka} and the Rosenbluth separation analysis of~\cite{Carlson:2001mp}.}
  \label{fig:he4_responses}
\end{figure}

In Fig.~\ref{fig:test_xsection}, we present the ANN predictions for four different nuclei and kinematics belonging to the test datasets.  Despite never encountering these kinematics before, the ANN is capable of capturing all reaction mechanisms, including elastic and quasi-elastic scattering, as well as the deep inelastic scattering region. The spread of the predictions is consistent with the size of experimental errors. For $^4$He and $^{12}$C, where a large amount of data is available for training, the results exhibit very good agreement with experimental data, with notably smaller error bars. Conversely, the uncertainty band is notably wider for $^{16}$O and $^{40}$Ca, primarily due to the scarcity of datasets for these two nuclei. In the case of $^{16}$O, where experimental data is limited, the ANN greatly benefits from information gathered on different nuclei with similar kinematics. We note that this capability arises from training the ANN simultaneously on various nuclei. In the Supplemental Material, we provide extensive ANN predictions for the entire test dataset. The excellent agreement between ANN predictions and experimental data corroborates the accuracy of the chosen architecture as well as the reliability of the Bayesian training.

As a second step of our analysis, we train the ANN on all the available experimental data to predict the responses of $^4$He, $^6$Li, $^{12}$C, $^{16}$O and $^{40}$Ca and compare them with Rosenbluth separation analysis found in the literature.

The ANN response functions of $^4$He at $q=400$ MeV/c, shown in Fig.\ref{fig:he4_responses}, are in remarkably good agreement with previous experimental extractions reported in~\cite{Carlson:2001mp}. At low energy transfers, the uncertainties are large due to low-lying nuclear states. The fine details of this part of the spectrum have not been accurately learned by the ANN owing to insufficient data and the fact that low-energy transitions strongly depend on the specific nucleus. We note that some relatively minor differences with Rosenbluth-separation analyses are visible in the tails of the quasi-elastic peak in both the longitudinal and transverse channels. We explicitly checked that ANN responses agree well with experimental data also for $q=300$, $500$, and $600$ MeV/c --- see the Supplemental Material for the corresponding figures. In general, the ANN yields smaller uncertainties for $R_T$ than for $R_L$. Consistent with what observed at $q=400$ MeV/c, the responses below $\omega=50$ MeV often tend to be unstable, leading to large uncertainties.

In Figure~\ref{fig:he4_responses}, the ANN longitudinal and transverse response functions are also compared with ab initio GFMC and LIT-CC calculations. The GFMC uses the highly realistic phenomenological Argonne v${18}$~\citep{Wiringa:1994wb} + Illinois 7~\citep{Pieper:2008rui} (AV18+IL7) Hamiltonian, which reproduces the spectrum of $A \leq 12$ nuclei with percent-level accuracy. The electromagnetic transition operator, largely consistent with the Hamiltonian, comprises one- and two-body terms and are derived within the so-called standard nuclear physics approach~\citep{Shen:2012xz}. On the other hand, the LIT-CC calculations are based on a chiral effective field theory Hamiltonian that includes terms up to next-to-next-to-leading order, without explicit $\Delta$ degrees of freedom, referred to as NNLO$_{\rm sat}$~\citep{Ekstrom:2015rta}. This interaction Hamiltonian includes two- and three-body terms optimized to simultaneously reproduce low-energy nucleon-nucleon scattering and selected nuclear structure data. Only one-body current contributions are retained in the transition operator.

\begin{figure}[!htb]
    \includegraphics[width=0.45\textwidth]{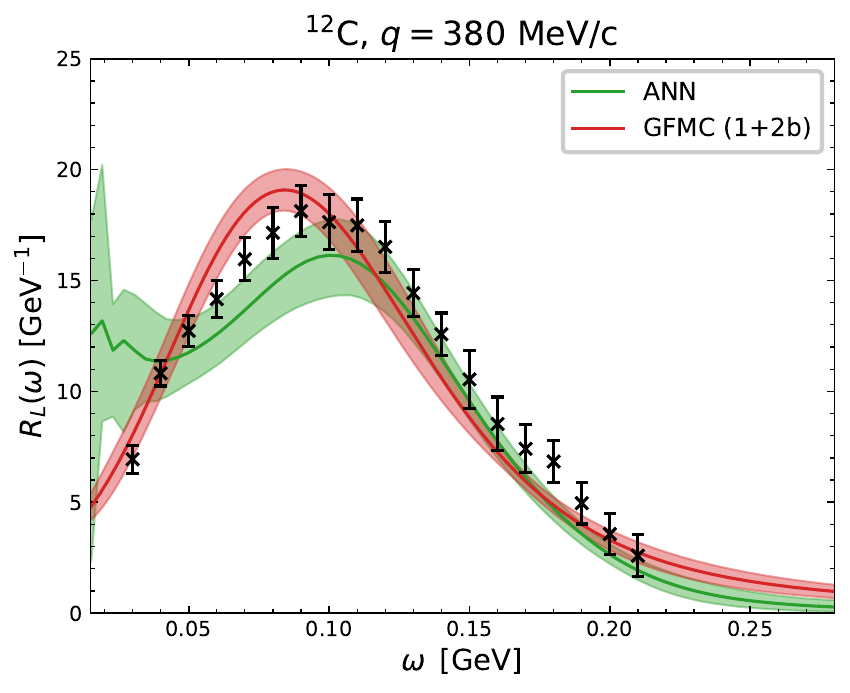}
\includegraphics[width=0.45\textwidth]{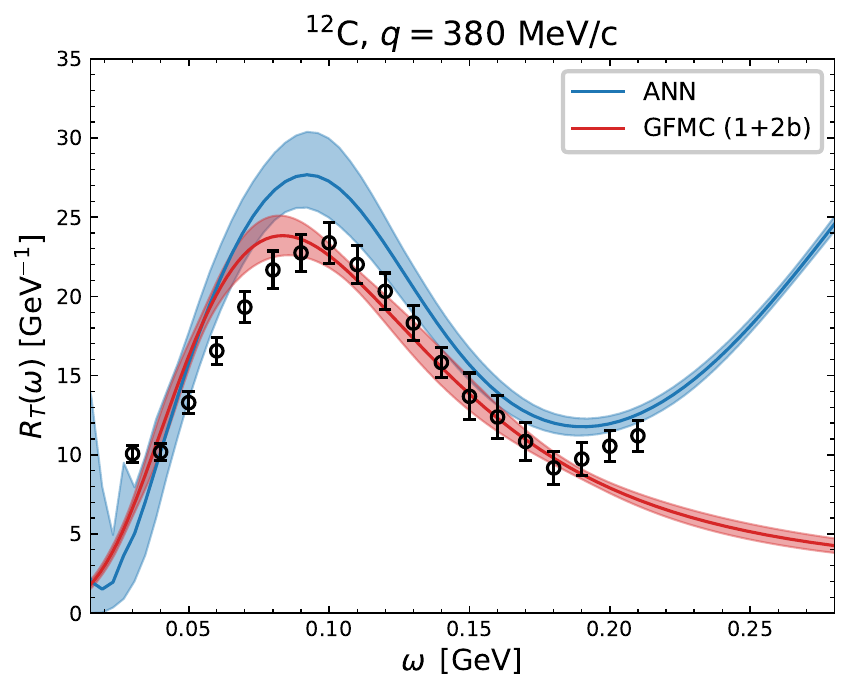}
  \caption{Electromagnetic responses on $^{12}$C for $q=380$ MeV/c. The ANN extractions are compared with theoretical calculations~\citep{Lovato:2016gkq} and the Rosenbluth-separations analysis of~\cite{Jourdan:1996np}}
  \label{fig:c12_responses}
\end{figure}

In the longitudinal channel, we observe remarkable agreement between the two ab initio methods, the ANN response functions, and the Rosenbluth separation analysis. However, in the transverse channel, the GFMC calculations exhibit an excess strength compared to the LIT-CC calculations, driven by two-body current contributions. In particular, it has been observed that this enhancement is primarily due to the interference between one- and two-body terms leading to final states with only one nucleon in the continuum~\citep{Fabrocini:1996bu,Benhar:2015ula,Franco-Munoz:2023zoa,Lovato:2023khk}. The discrepancies between the GFMC results and the ANN predictions in this channel might be attributed to relativistic corrections in the currents, which emerge at higher orders in the expansion compared to the longitudinal case~\citep{Rocco:2016ejr,Rocco:2018tes}, and have not been accounted for in the present work. It is also reassuring that the GFMC underestimates data on the right side of the quasi-elastic peak, as this allows for the accommodation of strength that is very likely to leak from the $\Delta$ region.
 
Our results for $^{12}$C at $q=380$ MeV/c in Figure~\ref{fig:c12_responses} closely match the Rosenbluth separation performed in~\cite{Jourdan:1996np} for the $R_L$ response. However, they predict more strength in the transverse channel; a similar trend is observed at $q=300$ MeV/c, as shown in the Supplemental Material. Due to the fact that some of the inclusive cross-section kinematics include elastic and inelastic transitions to low-lying excited states, the low-energy part of the spectrum exhibits larger uncertainties. The GFMC calculations of Ref.~\cite{Lovato:2016gkq} show good agreement with the ANN responses, although some differences are visible in the low $\omega$ region in the longitudinal channel. The GFMC and Rosenbluth separation match perfectly in that region. It is important to note that the contributions from elastic and low-lying inelastic transitions are explicitly removed from the GFMC responses and the Rosenbluth analysis, while they are present in the ANN curves. In the transverse channel, the ANN transverse response appears to be above both the Rosenbluth analysis and GFMC calculations at energies larger than the quasielastic peak. As discussed earlier for the transverse response of $^4$He, this behavior is reassuring as it leaves room for pion production in the $\Delta$ peak, which is not included in the GFMC and will produce strength in that region. 

\begin{figure}[!b]
    \includegraphics[width=0.45\textwidth]{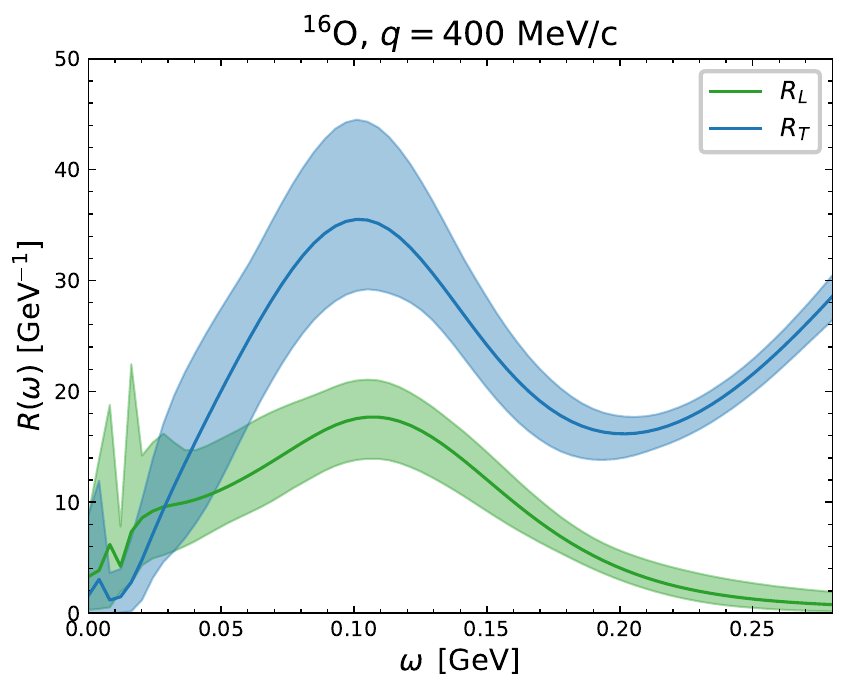}
  \caption{Electromagnetic responses on $^{16}$O for $q=400$ MeV/c. }
  \label{fig:o16_responses}
\end{figure}

Our ANN framework allows us to extract, for the first time, the longitudinal and transverse response functions of $^{16}$O, as shown in Figure~\ref{fig:o16_responses}. Owing to the scarcity of inclusive cross sections at different scattering angles, to the best of our knowledge, no Rosenbluth separation has been performed to extract the longitudinal and transverse responses of $^{16}$O. Since the ANN has been trained with limited $^{16}$O data, the Bayesian training automatically yields sizable uncertainties, much larger than for $^{4}$He and $^{12}$C, particularly at energy transfer up to the quasi-elastic peak region. For this nucleus, LIT-CC calculations will soon be carried out, whereas the computational cost renders this nucleus out of reach for GFMC. On the other hand, the auxiliary-field diffusion Monte Carlo method that has recently been applied to compute the Euclidean isoscalar density response of $A=16$~\citep{Gnech:2024qru} will soon be extended to accommodate electromagnetic longitudinal and transverse transition operators.

\begin{figure}[hbt]
    \includegraphics[width=0.45\textwidth]{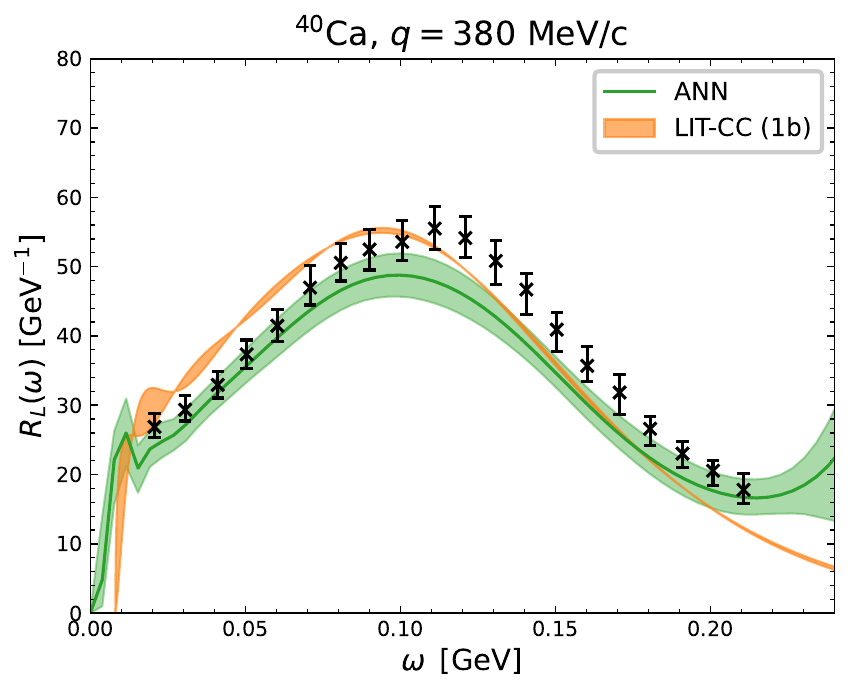}
\includegraphics[width=0.45\textwidth]{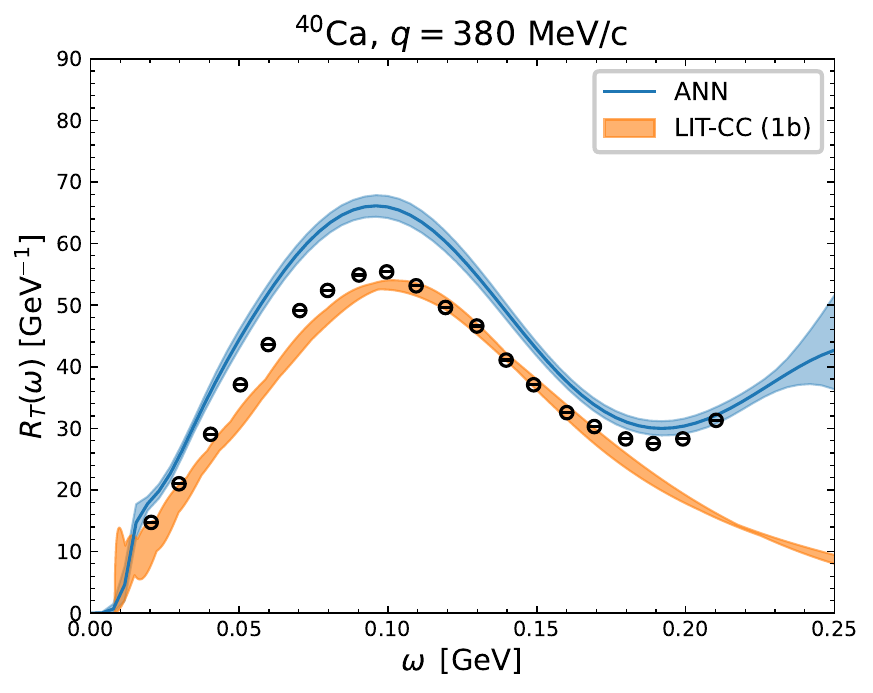}
  \caption{Electromagnetic responses on $^{40}$Ca for $q=380$ MeV/c. Our prediction compared with theoretical calculations~\cite{Sobczyk:2023sxh}. Data taken from~\cite{Jourdan:1996np}.}
  \label{fig:ca40_responses}
\end{figure}

The ANN results for $^{40}$Ca, shown in Fig.~\ref{fig:ca40_responses}, differ from the experimental data obtained from Rosenbluth separation, especially for $R_T$. It is interesting to note that the ANN uncertainties increase significantly in the high energy transfer region. This behavior, in contrast to what is observed in lighter nuclei, reflects the fact that there is little high energy-momentum transfer data available for $^{40}$Ca. Consequently, the ANN performs an extrapolation based on data available for other nuclei at high energies. The Bayesian training is fundamental in this regard, as it allows us to estimate the uncertainties associated with this extrapolation. The LIT-CC calculations for the longitudinal response are very close with both Rosenbluth-separation data and ANN predictions. In the transverse channel, it appears that including only the one-body current operator suffices to reproduce the Rosenbluth-separation data adequately, which is in contrast with what has been observed for $^4$He (and with the GFMC findings). However, it is noteworthy that the ANN predictions exhibit a $10-15\%$ enhancement compared to the experimental points. In this regard, we note that principally two experiments~\citep{Williamson:1997zz,Meziani:1984is} measured electron scattering on $^{40}$Ca and performed the Rosenbluth separation (there is one additional dataset~\cite{Whitney:1974hr}). As discussed in detail in the Supplemental Material, the results reported by these two analysis disagree substantially.

%The LIT-CC calculations for the longitudinal response closely match both the Rosenbluth-separation data and the ANN predictions. In the transverse channel, it appears that including only the one-body current operator suffices to adequately reproduce the Rosenbluth-separation data, which contrasts with what has been observed for $^4$He and with the GFMC findings. Importantly, the ANN predictions are about 15\% larger than Rosenbluth-separation, leaving room for two-body current contributions in the LIT-CC calculations. We note that principally two experiments measured electron scattering on $^{40}$Ca with sufficient precision to perform the Rosenbluth separation~\citep{Williamson:1997zz,Meziani:1984is} (there is one additional dataset~\cite{Whitney:1974hr}). As discussed in detail in the Supplemental Material, the results reported by these two analyses disagree substantially.

\section{Conclusions}
\label{sec:conclusions}

In this work, we performed the first extraction of electromagnetic longitudinal and transverse response functions using machine learning techniques for symmetric nuclei across a broad range of masses, $A=4-40$. A critical difference between our work and earlier studies~\citep{AlHammal:2023svo, Kowal:2023dcq}, which employed ANNs to directly model the $(e,e')$ inclusive scattering cross-sections, is that our ANN architecture outputs the longitudinal and transverse responses. These responses are then combined with the appropriate kinematic factors, which do not have to be learned, to obtain the inclusive $(e,e')$ cross section for a given incoming energy, scattering angle, and energy transfer. This procedure enables us to provide accurate predictions for $(e,e')$ inclusive cross sections on different nuclear targets, as well as to extract the longitudinal and transverse electromagnetic responses for various kinematics.

Our approach leverages Bayesian statistics to rigorously quantify the uncertainties in the ANN predictions. Specifically, we employ Hamiltonian Monte Carlo techniques to sample the posterior distribution of the ANN parameters, yielding a set of ANNs that are consistent with $(e,e')$ inclusive cross sections and fully account for the associated experimental errors. This Bayesian protocol also addresses epistemic uncertainties, automatically resulting in larger errors when extrapolating.

We obtained highly accurate results for $^4$He and $^{12}$C inclusive cross sections, benefiting from the availability of extensive training datasets. The algorithm successfully reproduces the test datasets for these cross sections across a wide range of energies, encompassing various reaction mechanisms and degrees of freedom. The ANN also reproduces well the test datasets for the other nuclei we considered: $^6$Li, $^{16}$O, and $^{40}$Ca. However, the theoretical uncertainties are larger due to the fact that there are fewer experimental data available for these nuclei compared to $^4$He and $^{12}$C.

As a second step, we utilized the entire $(e,e')$ inclusive cross section dataset to perform the first ANN-based extraction of longitudinal and transverse electromagnetic response functions. The ANNs are, in general, in good agreement with previous Rosenbluth separation analyses found in the literature~\citep{Jourdan:1996np,Carlson:2001mp}. The availability of longitudinal and transverse responses enables us to make a direct comparison with the GFMC and LIT-CC ab-initio quantum many-body methods. We find that both the GFMC and LIT-CC reproduce the ANN responses well, in both the longitudinal and transverse channels. Notably, our ANN analysis of $^{40}$Ca suggests a potential underestimation of the $R_T$ strength in previous Rosenbluth-separation extractions, confirming a tension between the Saclay~\citep{Meziani:1984is} and Bates~\citep{Williamson:1997zz} data for $^{40}$Ca, and will likely resolve the tension with LIT-CC calculations. New experimental measurements would be extremely valuable to resolve this tension.

The approach presented in this work allows us to predict electromagnetic responses in scenarios where traditional methods fail due to the lack of data. A chief example is $^{16}$O, where no traditional Rosenbluth separation has been performed yet. The availability of additional inclusive electron-scattering data off $^{16}$O will further constrain the ANN, thereby reducing the error bands of our predictions. This will be particularly relevant for the T2K and Hyper-Kamiokande neutrino-oscillation experiments, which use water Cherenkov detectors. 

The promising results obtained for electron-nucleus scattering in this work pave the way for extending this framework to neutrino-nucleus scattering. Inclusive neutrino-nucleus cross sections can be decomposed into five response functions. The vector responses are essentially the same for both electron and neutrino cases; the differences stem from final state interactions due to the distinct charge of the nuclear final state in the case of charged-current transitions. To accurately determine the pure axial and axial-vector components, we plan to leverage near detector data from various nuclear targets, thereby providing reliable predictions for the far detector.

\section{Acknowledgments}
The present research is supported by the U.S. Department of Energy, Office of Science, Office of Nuclear Physics, under contracts DE-AC02-06CH11357 (A.~L.), by the DOE Early Career Research Program (A.~L.),2 by the Fermi Research Alliance, LLC under Contract No. DE-AC02-07CH11359 with the U.S. Department of Energy, Office of Science, Office of High Energy Physics (N.R.), by the SciDAC-5 NeuCol program (A.~L., N.~R.), by the European Union’s Horizon 2020 research and innovation programme under the Marie Skłodowska-Curie grant agreement No.~101026014 (J.~E.~S.) and by Deutsche Forschungsgemeinschaft (DFG) 
through the Cluster of Excellence ``Precision Physics, Fundamental
Interactions, and Structure of Matter" (PRISMA$^+$ EXC 2118/1) funded by the
DFG within the German Excellence Strategy (Project ID 39083149) (J.~E.~S.). This research used resources of the Argonne Leadership Computing Facility, which is a DOE Office of Science User Facility supported under Contract DE-AC02-06CH11357 and the Laboratory Computing Resource Center of Argonne National Laboratory as well as the supercomputer MogonII at Johannes Gutenberg-Universit\"{a}t Mainz.

\bibliographystyle{elsarticle-harv}
\bibliography{biblio.bib}
\end{document}